\documentclass[journal]{journal}

\usepackage{algorithm}
\usepackage[noend]{algpseudocode}

\ifCLASSINFOpdf
   \usepackage[pdftex]{graphicx}
   \graphicspath{{/images/}}
\else
\fi
\usepackage[cmex10]{amsmath}

\begin{document}
%
\title{Boosting Method for Automated Feature Space Discovery in Supervised Quantum Machine Learning Models}
%
%
%

\author{Vladimir Rastunkov,~\IEEEmembership{IBM Quantum, Industry \& Technical Services,} \\
Jae-Eun~Park,~\IEEEmembership{IBM Quantum, Industry \& Technical Services,} \\
Abhijit~Mitra,~\IEEEmembership{IBM Quantum, Industry \& Technical Services,} \\
Brian~Quanz,~\IEEEmembership{IBM T.J. Watson Research Center, Yorktown Heights,} \\
Steve~Wood,~\IEEEmembership{IBM T.J. Watson Research Center, Yorktown Heights,} \\
Christopher~Codella,~\IEEEmembership{IBM Quantum, Industry \& Technical Services,} \\
Heather~Higgins,~\IEEEmembership{IBM Quantum, Industry \& Technical Services,} \\
Joseph~Broz,~\IEEEmembership{IBM Quantum, Industry \& Technical Services}
}

\maketitle
\thispagestyle{empty}

\begin{abstract}
Quantum Support Vector Machines (QSVM) have become an important tool in research and applications of quantum kernel methods. In this work we propose a boosting approach for building ensembles of QSVM models and assess performance improvement across multiple datasets. This approach is derived from the best ensemble building practices that worked well in traditional machine learning and thus should push the limits of quantum model performance even further. We find that in some cases, a single QSVM model with tuned hyperparameters is sufficient to simulate the data, while in others - an ensemble of QSVMs that are forced to do exploration of the feature space via proposed method is beneficial.
\end{abstract}


%
\IEEEpeerreviewmaketitle

\section{Introduction}
%
%
%
%


\IEEEPARstart{D}{ecision} support systems across multiple industries rely on heuristic approaches using models trained on historical data. For example, credit risk, propensity, attrition, fraud, and hospital readmission risk models classify data into two classes given a set of input features. The goal is to train analytical models that give the most accurate predictions, retain stable performance over time and utilize fewer features possible by efficiently extracting information from available data. Generally, the better the model can describe complex non-linear interactions between features, the higher its performance and ability to generalize the data will be. Conventional, non-quantum, machine learning models with higher performance are often achieved through ensembles of simpler learners. See, for example \cite{Freund1996machine, Breiman2001random, Hastie2017elements}.

Havlicek, et al. \cite{havlivcek2019supervised} implemented a quantum support vector machine classifier (QSVM) on a superconducting processor. Originally proposed in \cite{rebentrost2014quantum}, QSVM exploits a high-dimensional quantum Hilbert space to obtain an enhanced solution. This enhancement can be achieved through controlled entanglement and interference, which is inaccessible for classical support vector machines. However, superior performance of QSVM or other quantum approaches compared to traditional machine learning models is yet to be demonstrated on a practical dataset. Park et al. \cite{Park2020practical} demonstrated improvements to QSVM compared to classical SVM by using parameterized shallow unitary transformations for feature maps with rotation and regularization. Wu et al. \cite{wu2021application} provided benchmarks comparing performance of QSVM built using a simulator and physical hardware with classical SVM and \textit{xgboost}. Those benchmarks built using three different platforms, IBM Quantum, Google Tensorflow Quantum and Amazon Bracket, indicated similar performance of QSVM compared to its classical counterparts on a practical dataset. Another recent paper by Glick et al. \cite{glick2021covariant} discusses a class of covariant kernels and quantum advantage for problems where the data satisfies a group structure.

The idea of boosting quantum machine learning models was previously discussed by Neven et al. \cite{Neven2009training} in the context of adiabatic quantum computing implemented on D-Wave annealers, where authors used one level decision trees as weak classifiers. Papers by Schuld et al. and Abbas et al. \cite{Schuld2017quantum, Abbas2020onquantum} discussed quantum ensembles of quantum classifiers primarily from a perspective of speedup due to parallel calculation.

In general, quantum machine learning (QML) models consist of data encoding into qubits, a variational quantum circuit with trainable parameters, a classical cost function and an optimization algorithm. Most of the QML models constructed this way are mathematically related to quantum kernel methods \cite{Schuld2021quantum}. Notably, initial state preparation and subsequent unitary transformation with input features is carried out through a circuit called feature map. Unlike other types of machine learning algorithms, the choice of initial feature map in quantum support vector machines (QSVM) could yield unique decision boundaries, making QSVMs with different feature maps independent from each other. This characteristic of QSVM is well suited for implementing boosting algorithms, however, given a large number of choices of feature maps the automation of feature map/model selection and training process is quite desirable. 

In addition to achieving better performance with quantum models, there is a need to make them more user-friendly by automating the model selection and training process. Currently, model architectures are often derived from well-known physical models, e.g. an Ising model and respective Hamiltonian were used for feature mapping. Thus, an automated procedure can augment some of the unnecessary complexity of the existing models for users without a physics background as well as assist in discovering new model architectures.

The approach presented in the current work is different from results discussed in the referenced sources in the following respects. First, we are focused on universal quantum computing with gates that can run on superconducting qubits as implemented, e.g. in IBM Quantum System One. Secondly, even though we consider shallow circuits for kernel functions, those tend to be stronger than typical weak classifiers discussed in the literature, which is mostly based on decision trees. Thirdly, we implement an automated model selection on every boosting step to choose from different topologies and thus explore wider feature and model spaces. This process can be used to search for alternatives for broadly used Ising-type models. Fourthly, our approach is not constrained to classification tasks, it can be equally applied to regression tasks. Lastly, the focus of many prior results was on the possibility of quantum speedup for classical procedures, whereas our focus is primarily on the development of models with higher performance. In this work we are applying our approach to the problem of binary classification and use classification accuracy on the test sample as a measure of performance. \newline

\noindent Our main contributions are:
\begin{itemize}
\item A new ensemble method for QSVM that enhances model performance, when the data is difficult to model for a single learner.
\item Hyperparameter optimization for QSVM.
\item Simulation on multiple datasets to ensure stability of results.
\end{itemize}

\section{Boosting Method for QSVMs}

\subsection{Data and Data Encoding}

In this work we consider classical examples of generated data, such as moons, circles and XOR. This allows creation of many different datasets to accumulate statistics of model performance.

Following the best practice, the data is split into training, validation and testing datasets. A validation dataset is used for hyperparameter tuning in the process of grid search for the best model on every step of the boosting procedure. A testing dataset is completely hidden from the training and is used to compare different models.

\begin{figure*}[ht] 
	\centering
		\includegraphics[width=\textwidth]{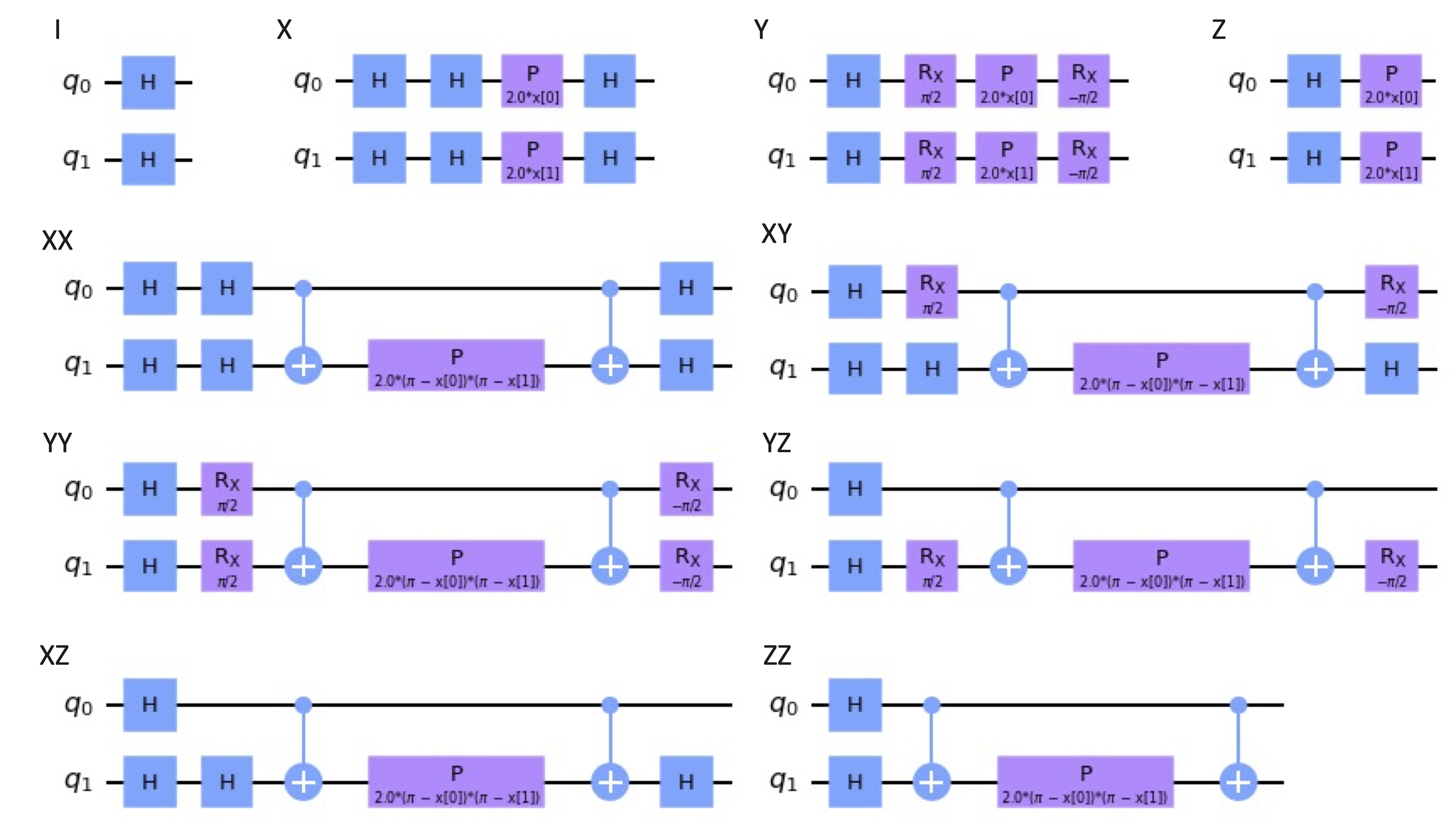}
	\caption{Set of feature maps for grid search.}
	\label{fig:feature_maps}
\end{figure*}

Following \cite{havlivcek2019supervised} we define a feature map on $n$-qubits as 
\begin{equation}
  \label{eq:feature_map}
  {\mathcal{U}}_{\Phi}(\vec{x})=U_{\Phi(\vec{x})}H^{{\otimes}n}
\end{equation}
where
\begin{equation}
  \label{eq:feature_map_diag_gate}
U_{\Phi(\vec{x})}=\exp\left(i\sum_{S\subseteq [n]}
\phi_S(\vec{x})\prod_{i\in S} P_i\right)
\end{equation}
Here $H$ is the Hadamard gate and $P_i \in \{ I, X, Y, Z \}$. A set of feature maps that we can utilize for a grid search, when the data has two features is shown in Fig. \ref{fig:feature_maps}.

\subsection{Ensemble Structure}

The traditional \textit{AdaBoost} variant of boosting relies on weak learners, such as decision stumps, that are trained on every iteration \cite{Hastie2017elements}. For each subsequent iteration it emphasizes examples that were previously misclassified by calculating and assigning or updating their weights. The final prediction is calculated by weighted majority vote of classifiers. In this work we consider support vector machines on quantum kernels $k(\vec{x}_{i},\vec{x}_{j})=|\langle {\mathcal{U}}_{\Phi}(\vec{x}_{i})|{\mathcal{U}}_{\Phi}(\vec{x}_{j})|^{2}$ that we call Quantum Support Vector Machines (QSVM). QSVM is not a weak learner, so we modify the boosting method as shown in Algorithm \ref{alg:bqsvm}.

\def\algbackskip{\hskip-\ALG@thistlm}
    \begin{algorithm*}
    \caption{Boosted QSVM classifier.}\label{alg:bqsvm}
    \hspace*{\algorithmicindent} \textbf{Input} {${X}_{train}, {y}_{train}, {X}_{val}, {y}_{val}, {y}_{train, i} \in \{0,1\}, {y}_{val, i} \in \{0,1\},$ grid parameters for QSVM} \\
    \hspace*{\algorithmicindent} \textbf{Output} $G(x)$
    \begin{algorithmic}[1]
    \State Initialize ${w_i}=1,$ $\forall{i}$.
    \For {$m=1$ to $M$}
      \State {Perform grid search and select the best classifier $G_m(x)$ on $ ({X}_{train},{y}_{train},{X}_{val}, {y}_{val})$ taking into account exclusions from the grid and training weights ${w_i}$}
      \State {Check early stopping conditions for perfect and worse than random guessing classification.}
      \State {Exclude selected feature map from grid parameters for next iterations.}
      \State {Compute ${err}_{m} = \frac {\sum_{i=1}^{N} {w_i} \cdot I({y}_{train, i} \neq G_m({X}_{train,i}))}{\sum_{i=1}^{N} {w_i}}$ (\textit{estimator error})}
      \State {Compute ${\alpha}_{m}=\log ((1-{err}_{m})/{err}_{m})$ (\textit{estimator weight})}
      \State {Set ${w}_{i} \gets {w}_{i} \cdot \exp[{\alpha}_{m} \cdot I({y}_{train, i} \neq G_m({X}_{train,i}))]$}
    \EndFor
    \State {Output $G(x) = \sum_{m=1}^{M} ({\alpha}_{m} G_m(x)) / \sum_{m=1}^{M} ({\alpha}_{m})$}
    \end{algorithmic}
    \end{algorithm*}

In the beginning the algorithm receives training and validation datasets as well as grid search parameters. In this work we consider the following parameters: Pauli feature map set as shown in Fig. \ref{fig:feature_maps}, the Pauli rotation factor, which is a multiplier to the Pauli rotations (\textit{alpha}) and a regularization parameter (\textit{C}) for sklearn's support vector classifier (SVC). We vary \textit{alpha} in the interval $(0; 2]$, \textit{C} in $[1; 100]$. All examples are initially assigned a weight of $1$. Grid search uses a validation dataset to select the best model. After the best model is selected we check early stopping conditions: 
\begin{enumerate}
  \item Estimator is perfect, i.e. estimator error on the training dataset is $\le 0$.
  \item Estimator is as bad as random guessing or worse, i.e. estimator error is $\ge 0.5$ for binary classification or $\ge 1 - \frac{1}{{N}_{classes}}$ for multiclass.
  \item The maximum number of classifiers is reached.
\end{enumerate}

The feature map selected on the current iteration is excluded from the grid search for next iterations. This is important to force the model to explore a broader Hilbert space and, consequently, different decision boundaries by choosing other feature maps for the quantum kernel. Finally, the weights are updated as shown in Algorithm \ref{alg:bqsvm}. Once any stopping condition is satisfied then the final model object is returned. This object can be used to build predictions for new samples as a weighted majority vote of classifiers included in the model.

It is worthwhile to highlight differences of the approach presented here from more traditional \textit{boosting}: 1) we perform a grid search for the best model on each iteration of the algorithm, 2) we enforce exploration of different model architectures through parameter grid constrains. 

Identifying the optimal number of estimators and ensemble pruning is generally outside of the boosting method description and is up to the user. In this work we will choose the optimal number of estimators based on the minimum error on the validation sample.

\subsection{Numerical Simulation Results}

\begin{figure*}[ht]
	\centering
		\includegraphics[width=\textwidth]{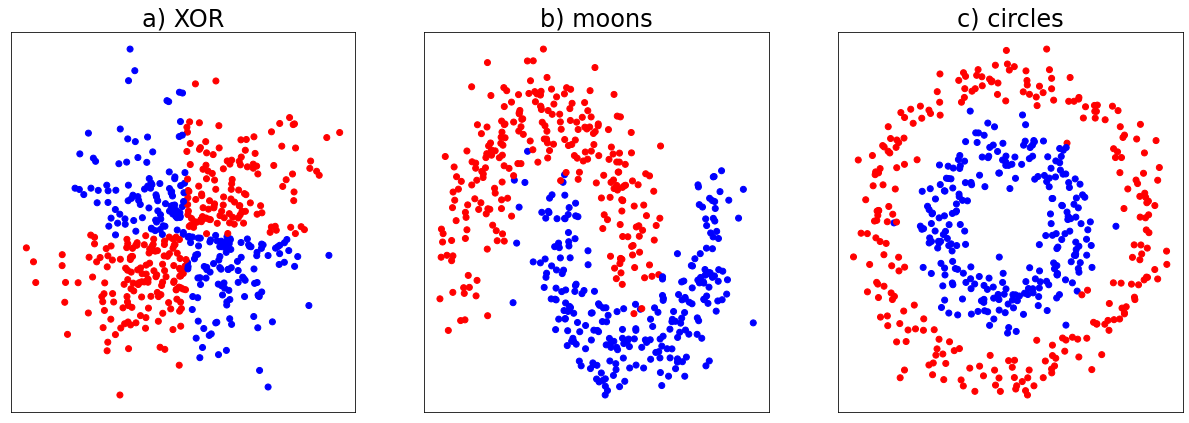}
	\caption{Different datasets used in experiments.}
	\label{fig:datasets}
\end{figure*}

First, we run experiments on simulated data created with functions available in scikit-learn (see Fig. \ref{fig:datasets}). This allows us to create a number of statistically independent datasets and obtain averaged performance metrics. In this study we chose to generate 50 datasets of each kind: XOR, moons and circles.

Each dataset has 150 observations for training, validation and testing split equally between those subsets. We train a boosted QSVM as described above for each dataset. For comparison, we also train an \textit{SVM} and \textit{xgboost}. The parameter grid for the \textit{SVM} includes RBF and linear kernels, regularization \textit{C} ranging from 0.1 to 100, gamma parameter for RBF kernel ranging from 0.0001 to 10. The parameter grid for \textit{xgboost} was constructed following \cite{wade2020handson}.

\begin{figure*}[ht]
	\centering
		\includegraphics[width=\textwidth]{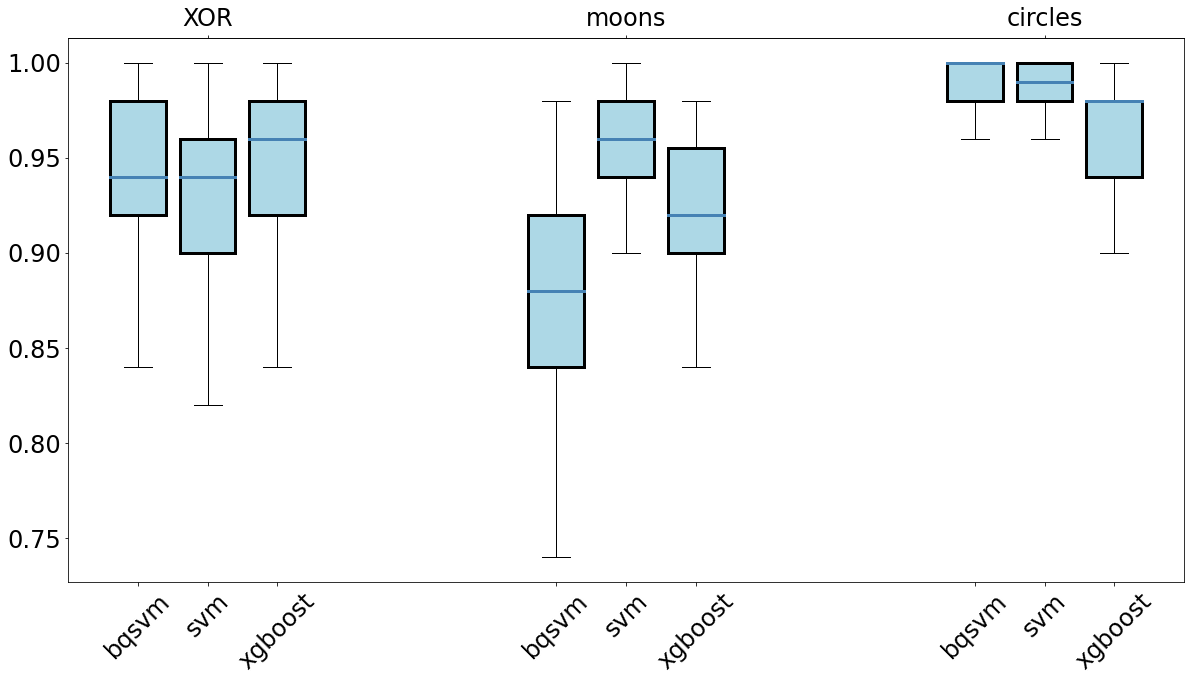}
	\caption{Model accuracy comparison box-plots. Lines show median accuracy on the test sample, boxes show the range between the lower and upper quartiles, and whiskers indicate "the range of the
  data" following Tukey's definition with $Q1 - 1.5 \cdot (Q3-Q1)$ and $Q3 + 1.5 \cdot (Q3-Q1)$ for lower and upper whiskers, respectively.}
	\label{fig:accuracies}
\end{figure*}

The results are shown in Fig. \ref{fig:accuracies}. The performance on the XOR dataset seems comparable across the three models. Boosted QSVM struggles to achieve comparable performance on the moons dataset, but works best on the circles dataset with median at $100\%$ accuracy.

An interesting question is whether a Boosted QSVM actually benefited from the ensemble and if so then how much improvement did it provide. It turns out that only about $31\%$ of Boosted QSVM models contain more than 1 estimator in the ensemble. Table \ref{tbl:lengths} shows mean and maximum ensemble size by dataset. The more difficult the dataset for QSVM is, the larger the ensemble seems to be: more than 1 estimator is barely used for circles data, while 3.8 estimators on average are used for moons data.

\begin{table}[ht]
  \caption{Average and Maximum Model Ensemble Size Per Dataset}
  \label{tbl:lengths} 
\begin{center}
  \begin{tabular}{ |c|c|c| } 
  \hline
  Dataset & Mean & Max \\
  \hline
  XOR & 2.02 & 10 \\ 
  circles & 1.06 & 3 \\ 
  moons & 3.84 & 10 \\ 
  \hline
  \end{tabular}
\end{center}
\end{table}

We have also investigated whether there is a performance gain from having multiple classifiers. Table \ref{tbl:improvements} shows classification accuracy improvement from an ensemble of QSVM classifiers compared to a single QSVM. There is a small sample size for circles data, where even a single QSVM is doing well. There is an average of $4.2\%$ and $7.5\%$ classification accuracy improvement for XOR and moons. 

\begin{table}[ht]
  \caption{QSVM Accuracy Increase with Boosting}
  \label{tbl:improvements} 
\begin{center}
  \begin{tabular}{ |c|c|c|l| } 
  \hline
  \hfil Dataset & \hfil Mean & \hfil Max & \multicolumn{1}{|p{3.2cm}|}{\centering Number of ensembles with $>{2}$ learners (out of 50)}\\\hline
  XOR & $4.2\%$ & $16.0\%$ & {\hfil 36}\\ 
  circles & $2.0\%$ & $2.0\%$ & {\hfil 2}\\ 
  moons & $7.5\%$ & $14.0\%$ & {\hfil 24}\\ 
  \hline
  \end{tabular}
\end{center}
\end{table}

\section{Conclusions}
 
Data scientists across multiple industries continue to push limits in their search for the best-in-class machine learning model that would provide a competitive edge. Quantum machine learning holds a promise of even higher performance than classical due to enhanced feature spaces. The approach discussed here is derived and adapted from the best ensemble building practices that worked well in traditional machine learning and thus should push the limits of model performance even further. Examples discussed in this work show that boosted QSVM ensembles outperform single QSVMs that in some cases allows them to match accuracy of non-quantum models, and in other cases - even exceed it.

\ifCLASSOPTIONcaptionsoff
  \newpage
\fi



\bibliographystyle{IEEEtran}
\bibliography{references}

\begin{thebibliography}{10}
\providecommand{\url}[1]{#1}
\csname url@samestyle\endcsname
\providecommand{\newblock}{\relax}
\providecommand{\bibinfo}[2]{#2}
\providecommand{\BIBentrySTDinterwordspacing}{\spaceskip=0pt\relax}
\providecommand{\BIBentryALTinterwordstretchfactor}{4}
\providecommand{\BIBentryALTinterwordspacing}{\spaceskip=\fontdimen2\font plus
\BIBentryALTinterwordstretchfactor\fontdimen3\font minus
  \fontdimen4\font\relax}
\providecommand{\BIBforeignlanguage}[2]{{%
\expandafter\ifx\csname l@#1\endcsname\relax
\typeout{** WARNING: IEEEtran.bst: No hyphenation pattern has been}%
\typeout{** loaded for the language `#1'. Using the pattern for}%
\typeout{** the default language instead.}%
\else
\language=\csname l@#1\endcsname
\fi
#2}}
\providecommand{\BIBdecl}{\relax}
\BIBdecl

\bibitem{Freund1996machine}
Y.~Freund and R.~E. Schapire, ``Experiments with a new boosting algorithm,''
  \emph{Machine Learning: Proceedings of the Thirteenth International
  Conference}, 1996.

\bibitem{Breiman2001random}
L.~Breiman, ``Random forests,'' \emph{Machine Learning}, p. 5–32, 2001.

\bibitem{Hastie2017elements}
T.~Hastie, R.~Tibshirani, and J.~Friedman, \emph{The elements of statistical
  learning: data mining, inference and prediction}.\hskip 1em plus 0.5em minus
  0.4em\relax Springer, 2017.

\bibitem{havlivcek2019supervised}
V.~Havl{\'\i}{\v{c}}ek, A.~D. C{\'o}rcoles, K.~Temme, A.~W. Harrow, A.~Kandala,
  J.~M. Chow, and J.~M. Gambetta, ``Supervised learning with quantum-enhanced
  feature spaces,'' \emph{Nature}, vol. 567, no. 7747, pp. 209--212, 2019.

\bibitem{rebentrost2014quantum}
P.~Rebentrost, M.~Mohseni, and S.~Lloyd, ``Quantum support vector machine for
  big data classification,'' \emph{Physical review letters}, vol. 113, no.~13,
  p. 130503, 2014.

\bibitem{Park2020practical}
J.-E. Park, B.~Quanz, S.~Wood, H.~Higgins, and R.~Harishankar, ``Practical
  application improvement to quantum svm: theory to practice,''
  \emph{arXiv:2012.07725}, 2020.

\bibitem{wu2021application}
S.~L. Wu, S.~Sun, W.~Guan, C.~Zhou, J.~Chan, C.~L. Cheng, T.~Pham, Y.~Qian,
  A.~Z. Wang, R.~Zhang, M.~Livny, J.~Glick, P.~K. Barkoutsos, S.~Woerner,
  I.~Tavernelli, F.~Carminati, A.~D. Meglio, A.~C.~Y. Li, J.~Lykken,
  P.~Spentzouris, S.~Y.-C. Chen, S.~Yoo, and T.-C. Wei, ``Application of
  quantum machine learning using the quantum kernel algorithm on high energy
  physics analysis at the lhc,'' 2021.

\bibitem{glick2021covariant}
J.~R. Glick, T.~P. Gujarati, A.~D. Corcoles, Y.~Kim, A.~Kandala, J.~M.
  Gambetta, and K.~Temme, ``Covariant quantum kernels for data with group
  structure,'' 2021.

\bibitem{Neven2009training}
H.~Neven, V.~S. Denchev, G.~Rose, and W.~G. Macready, ``Training a large scale
  classifier with the quantum adiabatic algorithm,'' \emph{arXiv:0912.0779},
  2009.

\bibitem{Schuld2017quantum}
M.~Schuld and F.~Petruccione, ``Quantum ensembles of quantum classifiers,''
  \emph{Sci Rep}, vol.~8, no. 2772, 2018.

\bibitem{Abbas2020onquantum}
A.~Abbas, M.~Schuld, and F.~Petruccione, ``On quantum ensembles of quantum
  classifiers,'' \emph{arXiv:2001.10833}, 2020.

\bibitem{Schuld2021quantum}
M.~Schuld, ``Quantum machine learning models are kernel methods,''
  \emph{arXiv:2101.11020}, 2021.

\bibitem{wade2020handson}
C.~Wade, \emph{\BIBforeignlanguage{eng}{Hands-On Gradient Boosting with XGBoost
  and scikit-learn [electronic resource] / Wade, Corey.}}, 1st~ed.\hskip 1em
  plus 0.5em minus 0.4em\relax Packt Publishing, 2020.

\end{thebibliography}
\end{document}